\documentclass[aps,pre,twocolumn,amssymb,showpacs]{revtex4}
\usepackage{epsfig}

\begin{document}

\newcommand{\plotwidth}{\columnwidth}  

\title{Experimental synchronization of spatiotemporal chaos}
\author{R.~Neubecker, B.~G{\"u}tlich}
\affiliation{Institute of Applied Physics,
  Darmstadt University of Technology,
  Hochschulstr.~6, 64289~Darmstadt, Germany.
  email: bjoern.guetlich@physik.tu-darmstadt.de} 

\begin{abstract}
We report the first experimental evidence of a successful synchronization of spatiotemporal chaos. The experiments were performed on two unidirectionally coupled, nonlinear-optical systems of single-feedback type. The synchronization was investigated for different degrees of complexity of the spontaneous structures.
In all cases, the cross-correlation between the two system states was found to increase with the strength of the coupling. 
Numerical simulations yield comparable results and throw a light on the role of spatial inhomogeneities, which hamper a perfect synchronization. 
\end{abstract}


%

\maketitle

It is well-known that two coupled chaotic oscillators can synchronize \cite{pikovsky,boccaletti02a}.
The extension of this concept to spatially extended systems is a rather new topic.
Spontaneously forming complex and dynamic spatial structures (''spatiotemporal chaos'') play a role in numerous fields, from biology over chemistry to different areas of physics  \cite{hohwal}, e.g.\ fluid dynamics, solid state physics, or nonlinear optics \cite{optpatt}.
While as bridge to the temporal phenomena, often spatially discrete system like coupled oscillators are regarded, we will here focus on continuous systems.

To achieve synchronization between two separate systems, the system states must be communicated between the both. In most pattern forming systems, such as in fluiddynamical or chemical systems, the experimental realization of this communication is rather difficult. The spatial distribution of physical quantities, like a flow field or the concentration of chemical compounds, must be manipulated dynamically by injecting an external signal. Therefore, up to now only numerical simulations of coupled prototype models are performed, mostly in one spatial dimension only 
\cite{stcsynchro1,parlitz,discretesynchro1,nonident,garcia01a}.
Both uni- or bidirectionally coupling, and the synchronization of not completely identical systems \cite{nonident} are regarded. Following concepts of controlling-chaos, the coupling signal is sometimes derived from a comparison of both system states, e.g.\ is the difference of two quantities.
Since the transfer of the full spatial distribution of a system state is considered to be a problem, often the possibility to use only few discrete coupling channels is investigated \cite{parlitz,discretesynchro1}. 
In the present report instead, we regard the coupling of the full, space- and time-continuous state of spatially two-dimensional systems.

Light waves are ideal as carrier for the coupling signal, since they can carry almost arbitrary spatial and temporal profiles. 
Hence, it almost suggests itself to use nonlinear-optical pattern forming systems, where the light field is already a central physical quantity. 
Our experiment belongs to the class of the so-called {\em single-feedback systems}, which have become quite popular over the last years \cite{dalesfirth,singlefeedb,hyasia,arecchi00a,akhmanovoro}.

In our system, a reflective {\em Liquid Crystal Light Valve (LCLV)} is used as optical nonlinearity, providing a self-defocusing, saturable Kerr-type nonlinearity. 
Its large nonlinear sensitivity allows us to realize large aspect ratio patterns with moderate laser powers.
The LCLV consists of two thin layers, separated by a mirror and sandwiched between two transparent electrodes to which a low ac voltage is applied \cite{lclv}.
One layer is a photoconductor {\em (PC)}, which changes its conductivity according to the intensity profile $I_w(x,y)$ of an incident light wave.
This results in a space dependence of the voltage drop over the second layer, a liquid crystal {\em (LC)}, affecting its effective refractive index. 
A light wave, passing the LC layer ({\em read side}) and being internally reflected, consequently acquires a transverse phase shift profile, which is determined by the intensity profile at the PC ({\em write side}).

\begin{figure}
  \centerline{ \epsfxsize=\plotwidth \epsffile{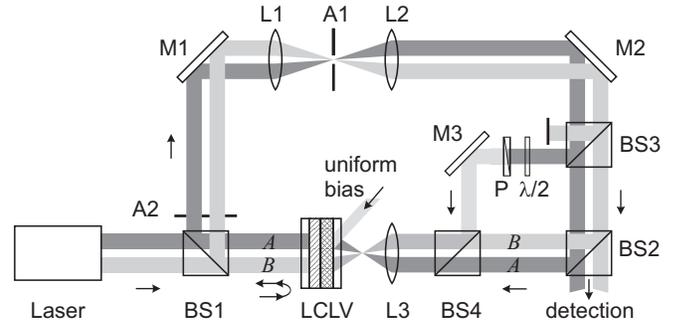}}  
  \caption{Schematic experimental setup. 
           For clarity, the pump beam is from the beginning already drawn as the
           two beams of the subsystems {\em A} and {\em B}. 
           Components for beam expansion, for detection and for the bias beam of 
           system $A$ are left out for simplicity.
           More details are given in the text.
  }
\label{sytusetup}
\end{figure}

The LCLV is put into an optical feedback loop: an almost plane pump wave is first phase-modulated and reflected by the LCLV read side. 
The modulated wave is then fed back to the intensity sensitive write side by means of beam splitters (BS), mirrors (M) and lenses (L), as indicated in Fig.~\ref{sytusetup}. 
During the free space propagation through the feedback loop, diffraction causes the phase profile to be transferred into an intensity modulation, such closing the feedback.
For detection purposes, a fraction of the feedback wave is coupled out with the beam splitter BS2. We use a digital camera to record an intensity distribution $I_w(x,y,t)$ equivalent to the one at the LCLV's write side. More details about this system and the theoretical modelling can be found in \cite{hyasia}.

\begin{figure}
  \centerline{ \epsfxsize=\plotwidth \epsffile{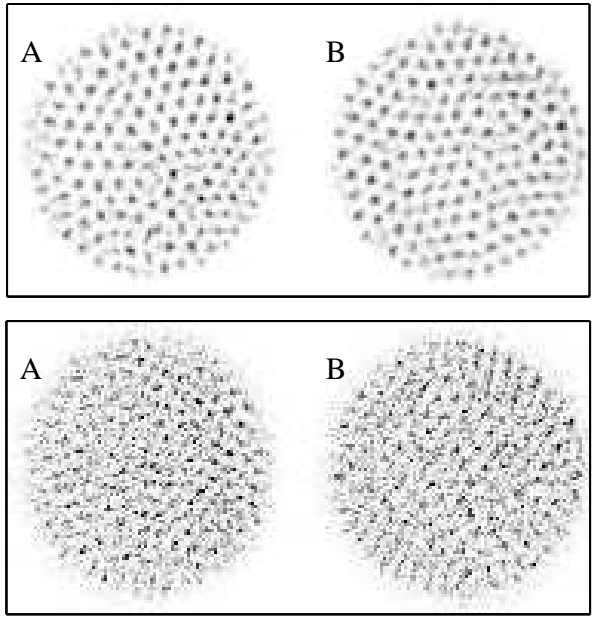}} 
  \caption{Snapshots of spatiotemporally chaotic states 
    without coupling ($\gamma= 0$) in inverse gray scale
    (dark corresponds to high intensity) at 1.5 times pattern forming threshold
    with closed low-pass filter (upper panel), 
    and with open low-pass filter (lower panel). 
    {\em A} and {\em B} denote master and slave system, respectively.
  }
\label{sytupatts1}
\end{figure}

Above a certain threshold of the pump intensity, the uniform, plane-wave solution becomes modulationally unstable with respect to a critical transverse wave number $k_c$. At higher respective thresholds, also higher order critical wave numbers become unstable.
Increasing the pump intensity from below threshold, first a stationary hexagonal pattern develops \cite{dalesfirth,hyasia}, and then the pattern becomes increasingly disordered and dynamic, finally turning into a very complex {\em spatiotemporally chaotic} state \cite{iapdyn,lclvdyn}. 
Typical snapshots of dynamic experimental structures are shown in Fig.~\ref{sytupatts1}. Just above threshold, remainders of perfect order are visible as hexagonal domains, the typical size of which shrinks with pump intensity. 

Our system includes a spatial low-pass filter in the feedback (aperture A1 in the Fourier plane between the lenses L1, L2).
By setting the low-pass cutoff just above the first critical wave number, the evolution of spatial chaos is impeded and the amount of disorder grows more smoothly with the pump intensity \cite{lclvdyn,leberre97a}. 
With an open low-pass filter instead, pronounced disorder sets in already closely above the primary threshold.
Hence, the closed low-pass allows us a finer adjustment of the degree of complexity.

Because of their versatility, LCLV feedback systems are frequently used to investigate spontaneous optical structures. In some configurations, the feedback wave is rotated or shifted laterally \cite{arecchi00a,akhmanovoro}. In the present version instead, the full rotational and translational symmetry is retained and diffraction represents the dominant spatial coupling mechanism.

We have recently shown, how this system can be synchronized by injecting perfect hexagonal patterns \cite{forclettpap}. 
The question addressed in this report is, whether such a synchronization can also be realized for spatiotemporal chaos. 
For this purpose, we regard two systems, one of which runs autonomously (master $A$), while the other (slave $B$) is exposed to the attenuated signal generated by the master system.
Since it is difficult to get two identical LCLVs, we divided the active area of a single LCLV into two independent subsystems by inserting a mask A2 with two circular holes (each diameter $D=4$~mm). 
Because of inhomogeneities of the LCLV, the two systems {\em A} and {\em B} cannot be expected to be perfectly identical.

In the feedback, a fraction of the light waves was extracted with the beam splitter BS3. 
While the signal from the slave system $B$ was blocked, the wave from the master $A$ was injected into $B$ with the beam splitter BS4. A halfwave-plate $\lambda/2$ and a polarizer P were used to attenuate the injected signal, i.e.\ to set the coupling factor $\gamma$. The perpendicular polarization behind the polarizer prevented interference effects. The recorded intensity distributions do not contain this coupling signal, because it was injected behind beam splitter BS2.

Due to the injected signal from $A$, the feedback intensity in $B$ changes to 
$ I_{wB}\rightarrow I_{wB}^\prime=I_{wB}+\gamma I_{wA}$.
Without further modifications, system $B$ would experience a total feedback power larger than system $A$. As consequence of the saturation of the nonlinearity, this would cause differences between systems $A$ and $B$.
In order to reduce this effect, a uniform bias beam was superimposed onto $A$, such that
$  I_{wA}\rightarrow I_{wA}^\prime= I_{wA}+I_{bias} $ with
$  I_{bias}= \gamma \langle I_{wA} \rangle$, where the brackets $\langle\cdot\rangle$ denote spatial averaging.
The seemingly simpler attenuation of $I_{wB}^\prime$ was hindered by technical reasons.

\begin{figure}
  \centerline{ \epsfxsize=\plotwidth \epsffile{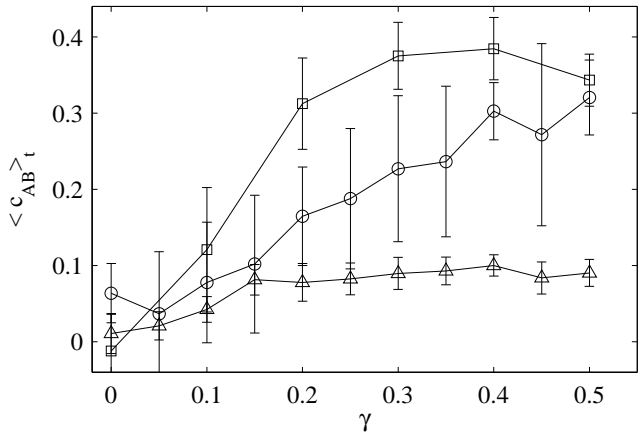}} 
  \caption{Time-averaged cross-correlation coefficient $\langle c_{AB}\rangle_t$, 
   plotted over the coupling strength $\gamma$. 
    $\square$:  pump intensity set to 1.5 times pattern forming threshold, 
    {\Large $\circ$}:  to 3 times threshold, 
    both for only one critical wave number active (closed low-pass); 
    $\vartriangle$: 1.5 times threshold with all critical wave numbers included 
    (open low-pass).
  }
\label{fpsynxcorr1}
\end{figure}

With the spatial low-pass filter cutoff set just above the critical wave number, the pump intensity $I_p$ was set to different values above the pattern forming threshold $I_{th}$. An additional measurement series was carried out with open low-pass and the pump at $I_p/I_{\rm th}=1.5$. In each particular measurement, the coupling strength was set to a fixed value and a temporal image sequence of up to 100~s duration was recorded. 
From each image, we computed a cross-correlation function between {\em A} and {\em B}
\begin{equation}
  C(\Delta x,\Delta y,t)= \frac{\langle \tilde I_{wA}(x,y,t)\cdot
                \tilde I_{wB}(x-\Delta x,y-\Delta y,t)\rangle}
               {\sqrt{\langle \tilde I_{wA}^2\rangle \langle \tilde I_{wB}^2\rangle}},
  \label{xcorrfct}
\end{equation}
where $I_{wA}$ and $I_{wB}$ are the experimentally recorded intensity distributions of master and slave system, respectively. The tildes stand for the deviation from the mean value $\tilde I= I-\langle I\rangle$. In order to exclude boundary effects, only the central parts of the active areas (85\% of the diameter) have been considered here.
As measure for synchronization, the correlation coefficient $c_{AB}(t)= C(0,0,t)$ was used.
Fig.~\ref{fpsynxcorr1} shows the time-averaged correlation coefficients, plotted over the coupling factor $\gamma$. The errorbars indicate the RMS value of the temporal variation during the recorded sequence. We have checked that the cross-correlation coefficients do not show temporal drifts, which confirms that the coupled systems were in their asymptotic states.

The increase of the cross-correlation coefficient $c_{AB}$ with the coupling strength $\gamma$ clearly substantiates that the two systems become synchronized by the coupling.
Because of the experimental imperfections and remaining differences between the systems $A$ and $B$, the correlation coefficient never reaches unity.
When the coupling parameter is larger than about $\gamma>0.5$, the correlation drops again.
This behavior can be assigned to the fact that the systems become increasingly different, because the bias intensity for the master system $A$ is an insufficient way to balance the unidirectional coupling.
Fig.\ref{fpsynxcorr1} also shows that the cross-correlation drops with increasing distance from threshold, meaning that the synchronization is harder to achieve for stronger disorder.

\begin{figure}
  \centerline{ \epsfxsize=\plotwidth \epsffile{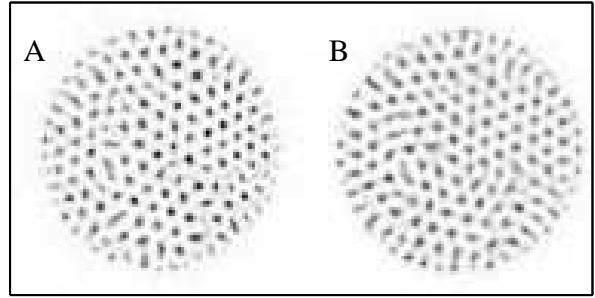}}  
  \caption{Snapshot of synchronized states with coupling factor $\gamma= 0.5$ at 
   1.5 times pattern forming threshold (low-pass filter closed).
  } 
\label{sytupatts2}
\end{figure}

Results very similar to the cross-correlation were obtained by computing the mutual information and the synchronization error \cite{boccaletti02a}, i.e.\ the normalized mean square difference between $I_{wA}$ and $I_{wB}$.
A snapshot of a synchronized state is presented in Fig.~\ref{sytupatts2}, for the pump intensity just above threshold and closed low-pass filter.
For weaker coupling, the correlations are smaller and difficult to be recognized by eye.

Without low-pass filtering, a larger number of critical wave number bands contributes to the pattern formation process, resulting in very complex and dynamic structures (cf.\ Fig.~\ref{sytupatts1}, lower panel).
In this case, the maximum achievable correlation is much lower and already saturates for $\gamma \geq 0.2$ (cf.\ Fig.~\ref{fpsynxcorr1}). 
The reduced correlation is probably the consequence of the increased sensitivity to perturbations. Without low-pass filtering, the system supports a much broader bandwidth and consequently high-frequency spatial noise, such as speckles, gain more influence. 
However, the increase of correlation at lower coupling strengths still gives evidence for a partial, coupling-induced synchronization.

In purely temporal systems,  a so-called {\em lag synchronization} can be found, where the synchronized system follows the master system with a certain time lag \cite{pikovsky,boccaletti02a}. In a spatially extended system this would correspond to a lateral shift (or rotation?) between the two patterns.
Such a shift would nota bene be connected to a symmetry breaking by the selection of a particular direction.
We find, however, that the dominant peak of the cross-correlation function Eq.~(\ref{xcorrfct}) does not change its location $(\Delta x=\Delta y=0)$. Hence, there is no lateral shift between the synchronized structures.

For comparison, numerical simulations of the full model equations were carried out. Details on the numerical algorithm are given in \cite{ampap}, we here used a grid with $128^2$ points. 
In one set of simulations, the situation was idealized, by using periodic boundary conditions and perfectly identical master and slave systems.
In the other set, circular boundaries were used. Moreover, we tried to include typical experimental imperfections, such as speckles and spatial variations of the nonlinear sensitivity. 
Since it is very difficult to quantify the experimental imperfections exactly, these simulations give only a qualitative picture.
To simulate speckles, we added white spatial noise to the pump beam (2\% of the total intensity in non-zero Fourier modes). The nonlinear sensitivity was given smooth spatial profiles, different for master and slave, with an variation of $\pm1\%$ (RMS) \cite{ampap}. 

\begin{figure}
  \centerline{ \epsfxsize=\plotwidth \epsffile{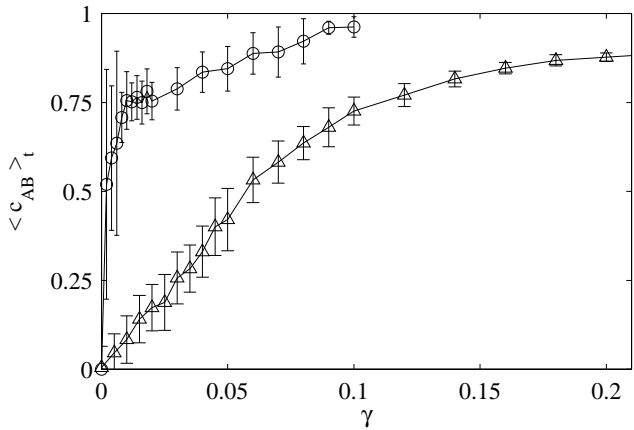}}  
  \caption{Cross-correlation coefficient $c_{AB}$ from numerical simulations, 
           for idealized systems at 7 times threshold ({\Large$\circ$}), 
           and with imperfections included at 4 times threshold ($\vartriangle$).
          }
\label{fpsytu2_kn}
\end{figure}

The resulting cross-correlation coefficients are shown in Fig.~\ref{fpsytu2_kn}. Even though the pump was set to 7 times threshold for the idealized case, the correlation grows quite fast with the coupling strength and reaches almost unity. Whereas when imperfections are included, the synchronization is still present, but the cross-correlation stays significantly smaller.

In conclusion, we have found a first experimental evidence for the synchronization of spatiotemporal chaos in a nonlinear-optical system, realized in a unidirectionally coupled master-slave configuration.
The dependence of the cross-correlation between master and slave state on the coupling strength clearly indicates synchronization.
Corresponding numerical simulations reveal how spatial inhomogeneities counteract synchronization.

The underlying mechanisms of this {\em synchorization} (from $\chi\omega\rho\iota o\nu$: place) of spatiotemporal disorder appear to be quite general; an observation in other nonlinear extended systems should be possible. Good candidates, besides other nonlinear optical systems, are photosensitive chemical reaction-diffusion systems, wich can both be detected and manipulated with light waves. A necessary translation between detection and controlling light or an amplification of the detection light can be performed by means of optically addressable spatial light modulators \cite{efronbook}.
The present work can also be considered as a step to secure parallel communication, as proposed in \cite{garcia01a}.

The authors acknowledge the support by T.~Tschudi and Jenoptik LOS GmbH, Jena.

\end{document}